\documentstyle[aps, epsfig,axodraw]{revtex}

\def\xslash#1{{\rlap{$#1$}/}}

\def\beq{\begin{equation}}
\def\eeq{\end{equation}}
\def\beqa{\begin{eqnarray}}
\def\eeqa{\end{eqnarray}}
\def\iar{\begin{array}{l}}
\def\ear{\end{array}}

\begin{document}
\draft
\title{Calculations of the Imaginary Part of Gauge Boson's 
Wave-function Renormalization Constants and the Modification of
hep-ph/0301090}
\author{Yong Zhou$^{a}$}
\address{$^a$Institute of Theoretical Physics, Chinese Academy 
of Sciences, P.O. Box 2735, Beijing 100080, China}
\date{Jun 5, 2003}
\maketitle

\begin{abstract}
We have calculated the imaginary part of W and Z gauge boson's
wave-function renormalization constants (wrc.) in the complex-pole
mass renormalization scheme and made atonement for the paper
hep-ph/0301090 where the imaginary part of gauge boson's
wrc. has been ignored in the second renormalization scheme in 
section 4. But the conclusion of this section in hep-ph/0301090
doesn't change.
\end{abstract}


In the complex-pole mass renormalization scheme\cite{c1},
the gauge boson's wave-function renormalization constant (wrc.)
shoul be complex number. If the gauge boson's self-energy 
function is writen as\cite{c2}
\beq 
  \begin{picture}(48,14)
    \Photon(0,5)(32,5){2}{6}
    \GCirc(39,5){7}{0.2}
    \Photon(46,5)(80,5){2}{6}
    \Text(5,13)[]{$V,\mu$}
    \Text(75,13)[]{$V,\nu$}
    \Text(5,-3)[]{$k$}
  \end{picture} \hspace{12mm} 
  \,=\, -i(g_{\mu\nu}-\frac{k_{\mu} k_{\nu}}{k^2})
  \Sigma^T_{VV}(k^2)-i\frac{k_{\mu} k_{\nu}}{k^2}\Sigma^L_{VV}(k^2) 
\eeq
then in the complex-pole mass renormalization scheme 
the form of the gauge boson's propagator in resonant region 
should be
\beq
  \frac{1}{p^2-M_0^2+\Sigma^T_{VV}(p^2)}\sim\frac{1}{m^2-M_0^2+
  \Sigma^T_{VV}(m^2)+(p^2-m^2)(1+\Sigma^{T\prime}_{VV}(m^2))}=
  \frac{(1+\Sigma^{T\prime}_{VV}(m^2))^{-1}}{p^2-m^2+B} 
\eeq
with $B=(m^2-M_0^2+\Sigma^T_{VV}(m^2))/(1+\Sigma^{T\prime}_{VV}(m^2))$
Thus we obtain the gauge boson's wrc.
\beq
  Z\,=\,(1+\Sigma^{T\prime}_{VV}(m^2))^{-1}\,
  \frac{one-loop}{}\hspace{-3mm}\rightarrow\,
  1-\frac{\partial}{\partial p^2}\Sigma^T_{VV}(m^2)
\eeq
In figure 1 we show the W gauge boson's self-energy diagrams
which have contribution to the $\xi_W$-dependent imaginary 
part of W gauge boson's wrc..
\begin{figure}[tbh]
\begin{center}
\epsfig{file=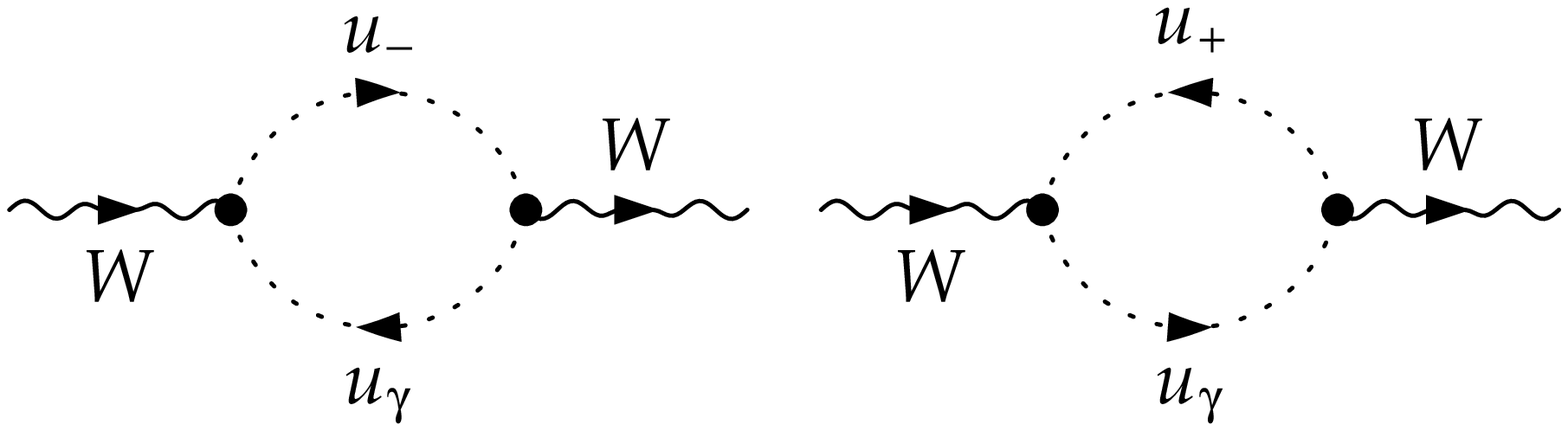,width=10cm}
\epsfig{file=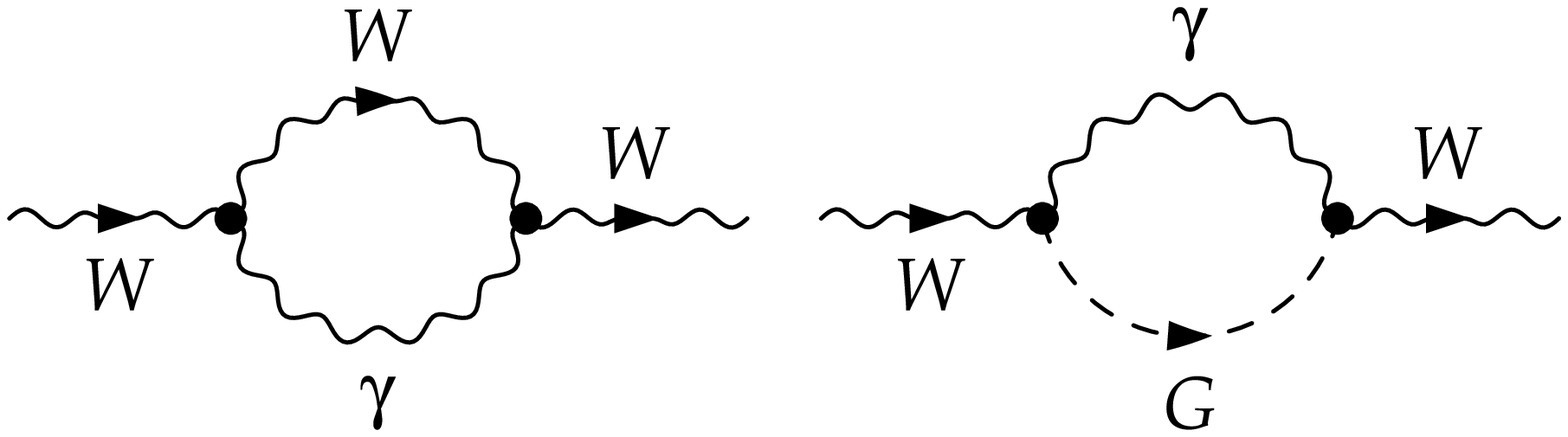,width=10cm}
\end{center} 
\caption{W boson's self-energy diagrams which have contribution
to the $\xi_W$-dependent imaginary part of W boson's wrc..}
\end{figure}
Based on Eq.(1),(3) and the {\em cutting rules}\cite{c3},
we have
\beq
  Im[Z_W]\,=\,-\frac{e^2(1-\xi_W)(\xi_W^2-2\xi_W-11)
  \theta[1-\xi_W]}{96\pi}
\eeq
In figure 2 we show the Z gauge boson's self-energy diagrams
which have contribution to the $\xi_W$-dependent imaginary 
part of Z gauge boson's wrc..
\begin{figure}[tbh]
\begin{center}
\epsfig{file=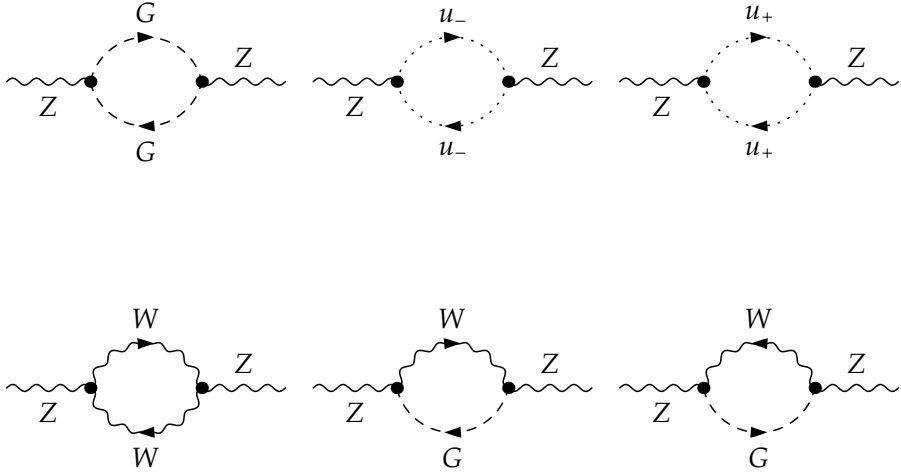,width=12cm}
\end{center} 
\caption{Z boson's self-energy diagrams which have contribution
to the $\xi_W$-dependent imaginary part of Z boson's wrc..}
\end{figure}
\hspace{-4mm}Based on Eq.(1),(3) and {\em cutting rules}, 
we also have
\beqa \iar
  \hspace{-4mm}Im[Z_Z]\,=\,-\frac{1}{96\pi c_W^2 s_W^2}
  e^2(1-4 c_W^2 \xi_W)^{3/2}\theta[1-4 c_W^2 \xi_W] 
  \\ \hspace{13mm}
  +\frac{1}{48\pi c_W^2}e^2((\xi_W-1)^2 c_W^4-2(\xi_W-5)c_W^2+1)
  \sqrt{(\xi_W-1)^2 c_W^4-2(\xi_W+1)c_W^2+1}\hspace{1mm}
  \theta[1/c_W-\sqrt{\xi_W}-1]
\ear \eeqa

Such imaginary part of gauge boson's wrc. will contribute to
the amplitudes of physical processes, e.g. 
$W^{-}\rightarrow e_i \bar{\nu}_i$ and 
$Z\rightarrow d_i\bar{d}_i$. These two processes have been
calculated in section 4 of hep-ph/0301090, but the   
imaginary part of gauge boson's wrc. has been ignored
in the complex-pole mass renormalization scheme.
So Eq.(57)-(60) of hep-ph/0301090 need to be modified. Eq.(58) and
(60) should be gotten rid of. Eq.(57) about the imaginary part of the
amplitude of $W^{-}\rightarrow e_i \bar{\nu}_i$ in complex-pole
mass renormalization scheme should be changed as follows
\beq
  Im[T_1^{\prime}]\,=\,A_L\frac{e^3(1-\xi_W)^3
  \theta[1-\xi_W]}{64\sqrt{2}\pi s_W}+B_L\frac{e^3(\xi_W-1)^3 
  m_{l,i}M_W\theta[1-\xi_W]}{32\sqrt{2}\pi s_W(M_W^2-m_{l,i}^2)}
\eeq 
It is also a gauge-dependent quantity. For the process
$Z\rightarrow d_i\bar{d}_i$, Eq.(59) is only the result of
the on-shell mass renormalization scheme. In the complex-pole
scheme, the right result is
\beqa \iar
  \hspace{-6mm}Im[{\cal M}(Z \rightarrow d_i \bar{d}_i) ]_{m_{d,i}
  \rightarrow 0} \,=\, A_L[-\frac{1}{576\pi c_W^3 s_W}e^3
  (1-4 c_W^2 \xi_W)^{3/2}\theta[1-4 c_W^2 \xi_W]
  \\ \hspace{20mm}
  +\frac{1}{288\pi c_W^3}e^3 s_W ((\xi_W-1)^2 c_W^4-2(\xi_W-5)c_W^2+1)
  \\ \hspace{20mm}
  \times \sqrt{(\xi_W-1)^2 c_W^4-2(\xi_W+1)c_W^2+1}\,
  \theta[1/c_W-\sqrt{\xi_W}-1]] \\ \hspace{20mm}
  +A_R[-\frac{1}{576\pi c_W^3 s_W}e^3
  (1-4 c_W^2 \xi_W)^{3/2}\theta[1-4 c_W^2 \xi_W]
  \\ \hspace{20mm}
  +\frac{1}{288\pi c_W^3}e^3 s_W ((\xi_W-1)^2 c_W^4-2(\xi_W-5)c_W^2+1)
  \\ \hspace{20mm}
  \times \sqrt{(\xi_W-1)^2 c_W^4-2(\xi_W+1)c_W^2+1}\,
  \theta[1/c_W-\sqrt{\xi_W}-1]]
\ear \eeqa
where $A_L=\bar{d}_i {\xslash \epsilon}\gamma_L d_i,
A_R=\bar{d}_i {\xslash \epsilon}\gamma_R d_i$. It is also
gauge-dependent. At one-loop level all of these gauge-dependent
terms cannot contribute to the modulus square of the physical 
amplitude. But we haven't still known how to cancel out the
contributions of these gauge-dependent terms to the modulus square
of the physical amplitude at two loop level. It needs to be done
in the future.

Thus the conclusion of section 4 of hep-ph/0301090 is still alive.
That's: based on the present knowledge we cannot judge which scheme
is the better, on-shell mass renormalization shceme or complex-pole
mass renormalization shceme.

\vspace{6mm}
{\bf \Large Acknowledgments}
\vspace{2mm}

The author thanks professor Xiao-Yuan Li for the enlightened 
discussion with him.

\end{document}